# Vibe Coding and Web Application Security: A Twin-Prompt Study*

**Darko Andročec**
University of Zagreb Faculty of Organization and Informatics
Department of Information Systems Development
Pavlinska 2, 42000 Varaždin, Croatia
dandrocec@foi.unizg.hr

***Abstract.*** *Large language models increasingly generate complete web applications from natural-language prompts, raising the question of whether explicitly requesting security best practice improves the result. We study six functionally distinct web applications, each generated in two prompt variants that are identical except for an appended security-requirements section: a baseline (A) and a security-aware (B) variant. All twelve programs were produced by the same agentic coding assistant and the same model version in a single, non-iterative generation round, and were then analyzed with static, dependency, dynamic and manual techniques, yielding 75 confirmed findings out of 85 candidates. The security-aware variant produced fewer confirmed findings in every application (24 versus 51) and contained no Critical or High issues; the most severe finding was detected only by manual testing. Because the corpus is small and each variant was generated once, we report descriptive observations rather than statistically established effects, and position the work as a preliminary study whose pipeline is being scaled to multiple models and repeated runs.*

## 1 Introduction

The rapid adoption of LLM-based coding assistants makes it possible to generate runnable web applications from short natural-language descriptions, a practice colloquially termed “vibe coding”, a label popularised by Karpathy (2025). This delegates security-relevant decisions – authentication, input handling, access control, secret management – to a model that optimises primarily for producing working code. A practitioner can, however, add explicit security requirements to the prompt. Whether such security-aware prompting meaningfully changes the security posture of the generated code is an open and practically important question.

This paper reports an exploratory, multi-method study that uses a twin-prompt design: each application is generated twice from prompts that are identical except for an appended security-requirements section. Both variants are produced by the same assistant, the same model version and the same single-shot procedure, so that the prompt is the only intended difference between the members of a pair. The resulting code is examined with complementary static, dependency, dynamic and manual techniques, and the tool outputs are consolidated into a single verified dataset. We address four research questions: (RQ1) which classes of weakness appear, and how are they distributed across the OWASP Top 10 (2021)? (RQ2) how does the security-aware prompt (B) affect the number and severity of confirmed findings relative to the baseline (A)? (RQ3) to what extent do SAST, DAST and manual testing detect different findings? (RQ4) do confirmed findings differ between the two programming languages studied?

Our contributions are a reproducible twin-prompt corpus of twelve LLM-generated applications together with the complete analysis pipeline; a verified, de-duplicated findings dataset mapped to OWASP and CWE; and descriptive evidence on the effect of security-aware prompting, on the OWASP distribution of findings, on the complementarity of detection methods and on differences between languages. The scope is deliberately narrow: one assistant, one model version, one generation run per variant and six applications. The study is intended as a preliminary investigation that establishes and validates the pipeline, and whose design is currently being scaled to several models and repeated generation runs.

* This is the author’s accepted manuscript of a paper accepted for presentation at the 37th Central European Conference on Information and Intelligent Systems (CECIIS 2026), September 16–18, 2026, Varaždin, Croatia. The final version will be published in the CECIIS 2026 Conference Proceedings.

# 2 Background and Related Work

Secure web development is commonly framed against the OWASP Top 10 (OWASP Foundation, 2021) and catalogued at the level of individual weaknesses by MITRE's Common Weakness Enumeration (MITRE, 2021). Two families of tooling dominate practice. Static application security testing (SAST) inspects source code without executing it, whereas dynamic application security testing (DAST) probes a running instance through its exposed interfaces. Both have well-known blind spots: SAST over-reports pattern-level and configuration-style issues and cannot observe runtime behaviour, while DAST sees only the surface it can reach and crawl. Software composition analysis (SCA) covers a third, largely disjoint concern, namely known vulnerabilities in third-party dependencies. These complementary limitations are the standard argument for combining automated methods with manual testing, and they motivate the pipeline described in Section 3.3.

Controlled generation studies provided the first systematic evidence that LLM-generated code carries security weaknesses. Pearce et al. (2022) prompted GitHub Copilot across scenarios targeting high-risk CWEs and found roughly 40% of the generated programs to be vulnerable. Khoury et al. (2023) reported that only five of twenty-one ChatGPT-generated programs were secure, although the model could often improve them when the weakness was pointed out to it.

User studies give a more mixed picture. Perry et al. (2023) found that developers with access to an AI assistant wrote less secure code while believing the opposite. Sandoval et al. (2023) measured only a small security impact in a low-level C task, and Asare, Nagappan and Asokan (2023) found that Copilot reintroduced known vulnerabilities in about a third of cases but was not consistently worse than the human developers against whom it was compared.

A second strand of work turns such observations into reusable benchmarks. SecurityEval collects 130 Python prompts covering 75 CWE classes and demonstrates a semi-automated evaluation in which generated completions are screened by static analysers (Siddiq & Santos, 2022). BaxBench moves the unit of evaluation from the function to the application: it defines 392 backend generation tasks across fourteen frameworks and six languages, validates functionality with test suites and assesses security by executing end-to-end exploits, reporting that exploits succeeded against approximately half of the functionally correct programs produced by each evaluated model (Vero et al., 2025). SusVibes applies the same logic to agentic, real-world software-engineering tasks: a frontier coding agent produced functionally correct solutions in 61% of 200 tasks spanning 77 CWE classes but secure ones in only 10.5%, and appending vulnerability hints to the request did little to close the gap (Zhao et al., 2025). Large-scale vendor evaluation points in the same direction: across more than 100 models and 80 coding tasks, roughly 45% of generated samples introduced an OWASP Top 10 weakness, with the highest failure rate in Java (Veracode, 2025). Our corpus is not a benchmark and is far smaller; it is complementary in that it holds the task, the model and the environment fixed and varies only the prompt, which cross-model benchmark rankings do not isolate.

Closest to our question is work that treats the prompt itself as the intervention. Tony et al. (2025) systematically compared prompting techniques for secure code generation across several LLMs and observed a reduction in weaknesses, most consistently for iterative techniques in which the model critiques and revises its own output. At the lightweight end of the spectrum the effect disappears: a short vulnerability hint left the security of agent-generated code essentially unchanged (Zhao et al., 2025). Iteration is not a guaranteed remedy either, since five rounds of AI-assisted refinement increased critical vulnerabilities by 37.6% in a controlled experiment (Shukla, Joshi & Syed, 2025). Field evidence indicates that the problem reaches production code: an empirical study of code generated by Copilot and comparable assistants in GitHub projects found security weaknesses in 29.5% of the analysed Python snippets and 24.2% of the JavaScript ones (Fu et al., 2025), and the Vibe Security Radar project traces real CVEs back to AI-authored commits, recording a rising monthly count through early 2026 (Georgia Tech Systems Software & Security Lab, 2026). What is missing from this picture is a controlled comparison at the level of a whole, deployable application in which everything except the prompt is held constant and the resulting code is examined by more than one detection method. That is the gap this preliminary study addresses.

# 3 Study Design and Methodology

## 3.1 Application Corpus

The corpus comprises six functionally distinct web applications spanning three frameworks and two languages, each generated in two prompt variants, yielding twelve programs (Table 1). The applications were not drawn from an existing benchmark. They were specified so that, between them, they exercise the security-relevant surfaces that dominate the OWASP Top 10 in web software: session and token authentication, per-user resource access, file upload and path handling, server-side URL fetching, e-commerce business logic and administrative access control. The last column of Table 1 records the categories that each specification was designed to place under test; this mapping was fixed before generation and is independent of what the tools later reported.

This design follows the scenario logic of CWE-indexed corpora such as SecurityEval (Siddiq & Santos, 2022) and of application-level benchmarks such as BaxBench (Vero et al., 2025), but at a much smaller scale and with a different purpose: a benchmark is built to compare models on a fixed task set, whereas this corpus is built to compare two prompts on a fixed model. Because the tasks are bespoke, the absolute counts reported in Section 4 are not comparable with published benchmark scores. Aligning part of the corpus with a public benchmark task set, so that cross-study comparison becomes possible, is the first item of future work (Section 7).

Table 1. The six-application corpus (each generated as variant A and B).

| App | Name | Stack | Security-relevant surface (targeted OWASP categories) |
|---|---|---|---|
| APP1 | Blog | Python / Flask | Session authentication, user-generated content, own-post editing (A01, A03, A07) |
| APP2 | Task Management API | JavaScript / Express | Token authentication, per-user resources (A01, A07) |
| APP3 | File Upload | Python / FastAPI | File type/size handling, server-side path construction (A01, A04) |
| APP4 | Shop with Comments | JavaScript / Express | Order and pricing business logic, user comments (A03, A04) |
| APP5 | Contact + URL Preview | Python / FastAPI | Server-side fetching of user-supplied URLs, application logs (A10, A01) |
| APP6 | Admin Dashboard | Python / Flask | Privileged access control, administrative session handling (A01, A07) |

## 3.2 Twin-Prompt Generation

All code was generated with Claude Code, an agentic command-line coding assistant, running the model Claude Opus 4.8. All twelve generations used the same assistant version, the same model version and the assistant's default settings; no custom instructions, sampling override, retrieval source or additional tool was configured, and the assistant's file-writing capability was the only capability used. Each generation started in a fresh session with no memory of previous applications or of the other variant. The exact invocations and timestamps are recorded in the command log of the replication package.

Each application has a single specification prompt that describes the desired functionality, endpoints and data model in neutral terms and mentions security nowhere. The baseline prompt (variant A) is exactly this specification. The security-aware prompt (variant B) is byte-identical to it with one section appended. That section is application-independent: it is the same text for all six applications, it states general requirements derived from the OWASP Top 10 (2021) categories, and it deliberately contains no hint as to which weaknesses the application in question is likely to exhibit. Table 2 summarises its requirement items; the verbatim text of both prompts is in the replication package.

The construction is intended to represent what a practitioner can realistically do in one step — ask for secure code — rather than an optimised security prompt. We did not draft several candidate security sections and select the best-performing one, and no prompt tuning of any kind was carried out. The study therefore measures the effect of one particular, untuned security section, which is a lower bound rather than the best achievable effect of security-aware prompting.

The generation architecture is deliberately minimal. Each variant was produced in a single, non-iterative interaction: one prompt, one response, no follow-up turns, no self-critique, no judge model, no security feedback loop and no repair step. This is the weakest of the intervention architectures examined in the literature — Tony et al. (2025) find iterative self-critique to be the more effective family — and it was chosen because any multi-turn or judge-based architecture introduces additional variables between the members of a pair and would confound the effect of the prompt itself. Whether an evaluate-and-repair loop over the same corpus improves on the single-shot result is a separate question, left to the extended study.

Three further controls preserve the comparison. First, generated source code was frozen: no manual edit was made after generation, for any reason. Second, each program received an intentionally security-neutral Dockerfile that supplies only a runtime and applies no hardening; this file is byte-identical across the two variants of an application, so container configuration is held constant by construction and cannot contribute to any A–B difference. All twelve images were confirmed to build and to serve over HTTP. Third, functional parity between the members of a pair was checked against the specification: for every application both variants exposed the specified endpoints and implemented the specified user flows, and those flows were exercised during dynamic and manual testing. This is smoke-test level assurance rather than a shared functional test suite, and it is recorded as a limitation in Section 6. The security controls present in the B variants — for example bcrypt password hashing, CSRF tokens, JWT expiry, environment-sourced secrets, SSRF allow-listing and secure-cookie flags — are what the model produced in response to the appended section, not post-hoc modifications.

Table 2. Requirement items of the security-requirements section appended to every variant-B prompt (summarized; the verbatim prompt text is in the replication package).

| Requirement item (summarized) | Primary OWASP category |
|---|---|
| Authenticate every request and authorize access to each resource per user; deny by default | A01 |

| Requirement item (summarized) | Primary OWASP category |
|---|---|
| Protect state-changing endpoints against cross-site request forgery | A01 |
| Store passwords only as salted hashes using an accepted algorithm | A02 |
| Read all secrets and credentials from the environment; never hard-code them | A02 / A05 |
| Use parameterized queries; validate input and encode output | A03 |
| Restrict uploads by type, size and destination path; never trust client-supplied file names | A04 |
| Disable debug modes and verbose error output; return generic error messages | A05 |
| Set session cookies with HttpOnly, Secure and SameSite; give tokens a limited lifetime | A07 |
| Validate and allow-list outbound URLs before any server-side fetch | A10 |
| Follow OWASP Top 10 (2021) practice throughout the application | all |

## 3.3 Analysis Pipeline

Every program was analysed by four families of technique, selected because their blind spots differ (Table 3). The same command set was applied to both members of every pair, and all commands were logged with timestamps.

*Static analysis.* Source code was scanned with Bandit 1.9.4 for Python and with ESLint 9.39.4 together with eslint-plugin-security for JavaScript. Semgrep 1.163.0 was run on all programs with the rule packs p/default, p/security-audit, p/owasp-top-ten and p/secrets, with telemetry disabled.

*Dependency analysis.* Declared dependencies were audited with pip-audit 2.10.0 for the Python projects and with npm audit for the JavaScript projects.

*Dynamic analysis.* Each program was started in an isolated container on a dedicated network and probed with OWASP ZAP (official zaproxy/zap-stable image) in three modes: a baseline passive scan, a full active scan, and an API scan for the two applications that expose an OpenAPI specification. Nikto 2.1.6 was run against the same targets for server- and configuration-level checks.

*Targeted manual testing.* Finally, each program was tested by hand against hypotheses that the automated tools cannot formulate, because they depend on the intended semantics of the application. sqlmap 1.10.5 was used to confirm or refute injection candidates, and purpose-written scripts probed insecure direct object references, business-logic abuse (price and quantity tampering, last-admin lockout) and authentication or session weaknesses (token forgery, missing expiry, cookie flags). Every manual test was recorded together with the behaviour a secure implementation should exhibit, and its outcome was decided against that expectation.

Table 3. Complementarity of the four analysis approaches.

| Approach | Tools | What it can observe | Principal blind spot |
|---|---|---|---|
| SAST | Bandit, Semgrep, ESLint + security plugin | Insecure patterns in source: hard-coded secrets, unsafe APIs, debug flags, unparameterized queries | No runtime or deployment behavior; reports patterns that may be unreachable or mitigated elsewhere |
| SCA | pip-audit, npm audit | Declared dependencies with known CVEs | Silent on first-party code (no findings in this corpus) |
| DAST | OWASP ZAP (passive, active, API), Nikto | Behavior of the running instance: missing headers, cookie flags, server exposure, injection responses | Only the reachable and crawlable surface; no view of source; cannot know the intended access-control rules |
| Manual / targeted | sqlmap, purpose-written scripts | Semantics-dependent weaknesses needing an oracle: IDOR, price and quantity tampering, last-admin lockout, token forgery | Effort-intensive and non-exhaustive; limited to the hypotheses the analyst formulates |

## 3.4 Data Consolidation and Labelling

Tool outputs were processed in two stages. A transcription stage read the raw machine-readable reports and recorded one row per tool finding, copying tool-provided CWE and OWASP values where present and leaving them blank otherwise; this produced 249 raw rows. A consolidation stage normalised each row to a common vulnerability class and de-duplicated by the triple (application, variant, class), so that one weakness reported by several tools collapses into a single row listing every detecting tool. This produced 85 consolidated candidate findings.

The author then reviewed every consolidated finding and assigned its severity, its OWASP category and a status of confirmed or false positive. Ten were judged false positives, for example a ZAP SQL-injection alert refuted both by sqlmap and by parameterised-query source, and a Nikto wp-config alert raised against a non-PHP application. Model-assisted scripts performed parsing, normalisation, arithmetic and plotting only; every security judgement, severity rating and interpretation in this paper is the author's.

## 3.5 Descriptive Statistics and Reproducibility

Descriptive statistics were computed over the confirmed findings only, and the figure was rendered from the resulting tables. The coding assistant appears in this study in two disjoint roles: as the subject of study, where it generated the application code, and as a scripted assistant that wrote and ran the parsing, dataset-building, statistics and plotting scripts. In the second role it performed no security testing and made no analytical judgement. Because the sample is small — six applications in two variants, one generation run each — all reported figures are descriptive and no inferential test was applied. All scripts, raw reports, consolidated datasets and the command log are retained so that the pipeline can be inspected and re-executed.

# 4 Results

Across the twelve programs, 85 candidate findings were consolidated: 75 were confirmed and 10 judged false positives. Dependency analysis reported no known-vulnerable dependencies in any variant, so all findings below concern first-party code and its configuration.

## 4.1 OWASP Distribution and Severity (RQ1)

Confirmed findings concentrate in A05 Security Misconfiguration (48 of 75), followed by A01 Broken Access Control (11) and A07 Identification and Authentication Failures (6); Table 4 gives the full distribution. The A05 total is inflated by two configuration classes that are identical in both variants by the security-neutral container design (running as root; binding all interfaces) and by missing hardening headers, and should therefore be read as a property of the deployment posture rather than as a prompt-driven difference. By severity, the confirmed set comprises 2 Critical, 11 High, 16 Medium and 46 Low findings; their distribution across the two variants is shown in Fig. 1 and discussed in Section 4.2.

Table 4. Confirmed findings by OWASP Top 10 (2021) category.

| OWASP category | Count |
|---|---|
| A01:2021 Broken Access Control | 11 |
| A02:2021 Cryptographic Failures | 3 |
| A03:2021 Injection | 4 |
| A04:2021 Insecure Design | 2 |
| A05:2021 Security Misconfiguration | 48 |
| A07:2021 Identification & Authentication Failures | 6 |
| A10:2021 Server-Side Request Forgery | 1 |
| **TOTAL** | **75** |

## 4.2 Effect of Security-Aware Prompting (RQ2)

The security-aware variant produced fewer confirmed findings than its baseline twin in every application: 51 across the A variants against 24 across the B variants, a mean of 8.5 against 4.0 per application (Table 5). The B variants contain no Critical- or High-severity confirmed finding, whereas the A variants account for both Critical findings and all eleven High findings (Fig. 1). What remains in the B variants is predominantly low-severity configuration material.

Because part of that residue is constant by design, we checked how much of the A–B difference it could explain. The container- and header-level classes are identical in both members of every pair, since the Dockerfile is byte-identical and neither prompt mentions deployment hardening. They therefore contribute equally to the A total and to the B total, and removing them leaves the A–B difference of 27 findings unchanged: they inflate the absolute counts, and in particular the A05 and Low totals, but they can neither produce nor mask the observed difference. That difference is carried entirely by application-level classes: hard-coded secrets, forgeable or missing authentication, plaintext password storage, missing access control, path traversal and unrestricted server-side fetching.

Table 5. Confirmed findings per application by variant (A–B difference).

| App | Language | A | B | A–B |
|---|---|---|---|---|
| APP1 | Python | 10 | 4 | 6 |
| APP2 | JavaScript | 6 | 2 | 4 |
| APP3 | Python | 8 | 5 | 3 |
| APP4 | JavaScript | 7 | 3 | 4 |
| APP5 | Python | 9 | 5 | 4 |
| APP6 | Python | 11 | 5 | 6 |
| **TOTAL** | | **51** | **24** | **27** |
| **Mean per app** | | **8.5** | **4.0** | |

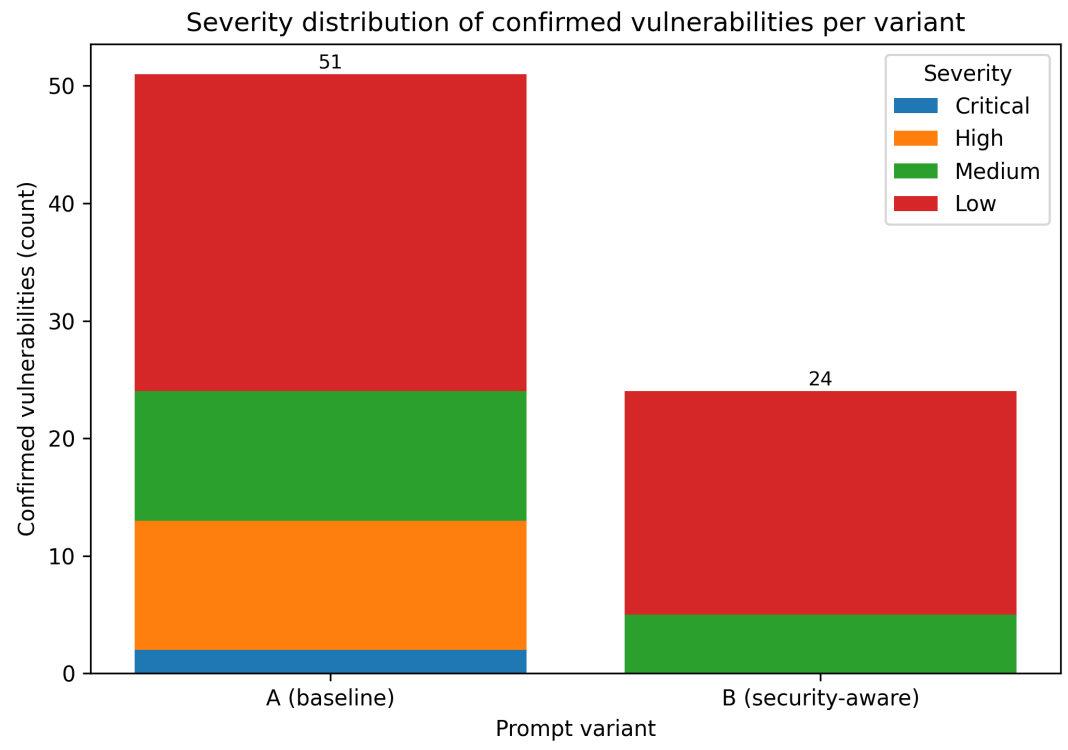

Figure 1. Severity distribution of confirmed findings per variant (A = baseline, B = security-aware).

### 4.3 Detection-Method Complementarity (RQ3)

The three detection approaches surfaced largely different findings. Of the 75 confirmed findings, 28 were found by SAST only, 24 by DAST only and 11 by manual testing only, while just 12 were corroborated by more than one method (Table 6). Counting any involvement, SAST contributed to 38 confirmed findings, DAST to 33 and manual testing to 16. The most severe issue — a JWT signed with a hard-coded secret, permitting forgery of arbitrary user tokens and account takeover in the baseline task API — was confirmed by manual testing and was not surfaced as an exploit by any automated scanner. This empirical overlap is the counterpart of the conceptual complementarity set out in Table 3.

Table 6. Confirmed findings by detection method.

| Detection method | Count |
|---|---|
| SAST only | 28 |
| DAST only | 24 |
| Manual only | 11 |
| Multiple methods | 12 |
| **TOTAL** | **75** |

### 4.4 Differences by Language (RQ4)

The four Python applications accounted for 57 confirmed findings (mean 14.25 per application) and the two JavaScript applications for 18 (mean 9.0). Because the corpus contains unequal numbers of applications per language and the applications differ in functionality, this comparison is confounded and should be read only as a description of the present corpus, not a language-level effect.

### 4.5 Notable Confirmed Vulnerabilities and False Positives

Among the baseline variants, testing confirmed a Critical JWT forgery (APP2) and a Critical forgeable administrator session (APP6), both caused by hard-coded secrets; a High server-side request forgery and a High unauthenticated admin-log exposure (APP5); a High path-traversal file write (APP3); High Flask debug mode in two applications; and plaintext password storage in three. Each corresponding B variant removed or mitigated the issue. Ten findings were judged false positives, notably a ZAP SQL-injection alert refuted by sqlmap and by parameterised-query source, and a Nikto wp-config alert on a non-PHP application. That the single most severe automated alert was a false positive, while the single most severe true issue was found only manually, illustrates why several methods and manual verification were used.

## 5 Discussion

Within this corpus, adding an explicit security-requirements section to the prompt was associated with fewer and less severe confirmed findings in every application, and with the elimination of all Critical and High confirmed issues. The security-aware variants did not become finding-free: what remained were predominantly low-severity configuration matters, part of them artefacts of the deliberately neutral container design that is held constant across variants. The pattern therefore suggests that security-aware prompting can move an application away from the most serious classes of weakness while leaving routine hardening to deployment configuration. It also offers a tentative contrast to benchmark evidence that lightweight vulnerability hints do not improve security (Zhao et al., 2025): here an explicit and fairly complete OWASP requirements section, rather than a brief nudge, accompanied the reduction in serious findings. The contrast is suggestive at best, since the corpus is small, the tasks differ from those benchmarks, and a single generation run per variant cannot separate the prompt effect from run-to-run variability.

The detection-method results reinforce a practical point: SAST, DAST and manual testing were largely complementary, with only a minority of findings seen by more than one method. Automated scanners produced both false negatives (missing the account-takeover-grade JWT forgery) and false positives (a High-severity SQL-injection alert that did not reproduce). Relying on any single method, and especially on a single automated scanner, would have materially misrepresented the security posture of these applications.

A third observation concerns the architecture of the intervention. The reduction reported here was obtained with the simplest possible arrangement — one prompt, one response, no review loop — which is also the arrangement that requires the least effort and cost from a practitioner. Whether it approaches what iterative self-critique achieves (Tony et al., 2025), or whether iteration would erode the gain as reported by Shukla,

Joshi and Syed (2025), cannot be answered from this design and is left to the extended study.

# 6 Threats to Validity

*Construct validity.* The construct of interest is the security posture of a generated application, and it is measured here by the number and severity of confirmed findings. That measure is imperfect in two respects. Severity ratings and OWASP/CWE mappings were assigned by a single analyst and are partly subjective, and the decision to de-duplicate by vulnerability class rather than by tool alert determines the counts, since a class reported on twenty endpoints counts once. To mitigate this, the raw pre-consolidation dataset and all scripts are retained, so that the counts can be re-derived under a different rule. Counting also treats a missing security header and an account-takeover-grade token forgery as one unit each, which is why severity is reported alongside every count.

*Internal validity.* Each variant was generated exactly once and LLM output is non-deterministic, so part of any A–B difference may be run-to-run variation rather than an effect of the prompt. This is the most serious limitation of the design and the reason no causal claim is made: the consistent direction of the difference across all six pairs is suggestive, but one observation per cell cannot separate the prompt effect from sampling noise. Repeated generation, with at least five runs per cell and reported dispersion, is the central change planned for the extended study. Labelling was performed by one author, without a second rater or a formal agreement measure. The security-neutral Dockerfile is shared across variants by design; as shown in Section 4.2 this suppresses container-level differences and cannot generate the observed difference, but it does inflate the absolute counts.

*External validity.* The corpus is small and bespoke: six applications, two languages, three frameworks, all generated by one assistant and one model version, from tasks written by the author rather than taken from a public benchmark. The results describe this corpus. They may not transfer to other models or assistants, to other languages and frameworks, to larger or legacy codebases, or to security prompts other than the one used here. The language comparison in Section 4.4 is additionally confounded by unequal group sizes and by differing application functionality. Functional parity between the members of a pair was established at smoke-test level rather than by a shared test suite, so residual functional differences within a pair cannot be excluded.

*Conclusion validity.* Because of the small sample we report descriptive figures only and apply no significance test; with six pairs and one observation per cell, no inferential procedure would carry meaningful power. The observed A–B differences should be read as a consistent pattern within this corpus, not as a statistically established effect.

# 7 Conclusion

We presented a multi-method, twin-prompt study of the security of LLM-generated web applications. Using static, dependency, dynamic and manual analysis over six applications generated in baseline and security-aware variants by a single model, we consolidated 85 candidate findings into 75 verified ones. Descriptively, security-aware prompting was associated with fewer confirmed findings in all six applications and with the removal of every Critical and High confirmed issue, while the detection methods proved strongly complementary and the most severe true finding was uncovered only by manual testing. These observations motivate the extended study now in preparation, which addresses the limitations above along four axes: repeated generation runs per cell, so that dispersion can be reported and inferential analysis becomes meaningful; several models and assistants, so that prompt effects can be separated from model characteristics and related to published baselines; alignment of part of the corpus with a public benchmark task set, to permit cross-study comparison; and a comparison of the single-shot security prompt with iterative and judge-based architectures. The application corpus, prompts, raw tool reports, consolidated datasets, statistics tables, figures, processing scripts and command log are available for inspection and re-execution in the public repository (https://github.com/dandrocec/vibe_coding_security).

# Acknowledgments

This research was funded by European Regional Development Fund – ERDF, under the project: S.A.M.I.R. – Sustainable AI-based Mobile Infrastructure for Renewable Charging and National Recovery and Resilience Plan 2021–2026, Institutional Research Projects under the project PiPi: Data infrastructure and augmented intelligence for sustainable development.